\documentclass[twocolumn,pra]{revtex4}
\usepackage{amssymb}
\usepackage{amsfonts}
\usepackage{amsmath}
\usepackage{graphicx}

\begin{document}

\title{Quantum non-Markovian processes break conditional past-future
independence}
\author{Adri\'{a}n A. Budini}
\affiliation{Consejo Nacional de Investigaciones Cient\'{\i}ficas y T\'{e}cnicas
(CONICET), Centro At\'{o}mico Bariloche, Avenida E. Bustillo Km 9.5, (8400)
Bariloche, Argentina, and Universidad Tecnol\'{o}gica Nacional (UTN-FRBA),
Fanny Newbery 111, (8400) Bariloche, Argentina}
\date{\today }

\begin{abstract}
For classical Markovian stochastic systems, past and future events become
statistically independent when conditioned to a given state at the present
time. Memory non-Markovian effects break this condition, inducing a
non-vanishing conditional past-future correlation. Here, this classical
memory indicator is extended to a quantum regime, which provides an
operational definition of quantum non-Markovianity based on a minimal set of
three time-ordered quantum system measurements and post-selection. The
detection of memory effects through the measurement scheme is univocally
related to departures from Born-Markov and white noise approximations in
quantum and classical environments respectively.
\end{abstract}

\maketitle



The definition of Markovianity and non-Markovianity in a quantum regime has
changed in time. Given that the physical essence of a classical (memoryless)
Markov approximation leads to a local-in-time evolution for a probability
density \cite{vanKampen}, accordingly a\textit{\ quantum Markovian regime}
was originally associated to local-in-time (non-unitary) density matrix
evolutions \cite{GardinerQ,breuerbook}. Hence, approximations that guarantee
this property, such as the well-known Born-Markov and white noise
approximations \cite{GardinerQ,breuerbook,cohen,milburn}, were related to
quantum Markovianity. The underlying assumptions in these approximations are
a weak system-environment coupling while the environment fluctuations define
the minor time-scale of the problem. Consequently, departure from these
physical conditions was associated to a \textit{quantum non-Markovian regime}
\cite{breuerbook,vega}.

In the last years the previous paradigm changed drastically. The more
general local-in-time evolutions that preserve the density matrix properties
(usually known as Lindblad equations) are established by the rigorous theory
of quantum dynamical semigroups \cite{alicki}. General behavioral properties
of the system propagator and different quantum information measures can be
established in this context. Thus, in last years quantum non-Markovianity
has been defined by departures from these \textquotedblleft canonical
behaviors\textquotedblright\ \cite{BreuerReview,plenioReview}. While strong
progress have been made on this basis \cite%
{BreuerReview,plenioReview,BreuerFirst,cirac,rivas,DarioSabrina,fisher,dario11,geometrical,mutual,fidelity,brasil,canonicalCresser,cresser,Acin,eternal,maximal}%
, some undesirable aspects have emerged. For example, in these novel
approaches dynamical departures from a Born-Markov approximation may be
included in a Markovian regime. This incongruence is present in almost all
proposals. On the other hand, in a incoherent limit, the classical notion of
Markovianity may not be recovered. Given that quantum systems are
intrinsically perturbed by measurement processes, a lack of an equivalent
operational (measurement-based) definition is also usual.

The aim of this work is to introduce an alternative approach to quantum
non-Markovianity that surpasses all previous drawbacks, which in turn is
consistent with the former\ (physical) notion of quantum Markovianity. The
proposal relies on post-selection techniques \cite{vaidman} and retrodicted
quantum measurements \cite{murch,molmer}, formalisms that allow inferring
the state of a quantum system in the past. Thus, the present approach brings
an active and fundamental area of research \cite%
{vaidman,murch,molmer,haroche,huard,naghi,decay,retro,wiseman,dressel,barnett}
into contact with the characterization of memory effects\ in open quantum
system dynamics.

A notable progress in the formulation of quantum memory indicators
consistent with classical non-Markovianity was introduced in Ref. \cite{modi}%
. Based on the usual definition of classical Markovianity in terms of
conditional probability distributions \cite{vanKampen} an operational based
\textquotedblleft process tensor\textquotedblright\ formalism defines
quantum non-Markovianity. The main theoretical component of the present
approach is similar but relies on an alternative and equivalent formulation
of classical Markovianity: the\textit{\ statistical independence of past and
future system events when conditioned to a given state at the present time} 
\cite{CoverTomas}. Hence, here a hierarchical set of conditional past-future
(CPF) correlations indicate departure from a classical Markovian regime. The
quantum extension of this alternative formulation leads to an operational
definition of quantum non-Markovianity based on a \textit{minimal} set of
three time-ordered successive measurements performed solely on the quantum
system. Post-selection introduces the conditional character of the quantum
measurement scheme. Furthermore, a non-vanishing CPF correlation, which has
the same meaning and (average) structure as in a classical regime, becomes a
univocal indicator of departures from Born-Markov and white noise
approximations in quantum and classical environments respectively.
Analytical solutions of relevant system-environment interaction models
support the formalism and conclusions.

\textit{Conditional past-future independence}: The observation of a
classical stochastic system at three successive times $t_{x}<t_{y}<t_{z}$
yields the outcomes $x\rightarrow y\rightarrow z$ (see Fig.~1). For a Markov
process, the joint probability distribution $P(z,y,x)$ of a particular
sequence can be written as $P(z,y,x)=P(z|y)P(y|x)P(x)$ \cite{vanKampen},
where $P(x)$ is the probability of the first event and, in general, $P(b|a)$
is the conditional probability of $b$ given $a.$ From here and Bayes rule,
the conditional probability $P(z,x|y)$ of future $(z)$ and past $(x)$\
events \textit{given} the present state $y$ is \cite{CoverTomas}%
\begin{equation}
P(z,x|y)=P(z|y)P(x|y).  \label{Indenpendence}
\end{equation}%
Thus, for a classical Markovian process past and future events become
statistically independent when conditioned to a given (fixed) intermediate
state. This property can be corroborated through a \textit{conditional
past-future correlation,} which is defined as%
\begin{equation}
C_{pf}\equiv \langle O_{z}O_{x}\rangle _{y}-\langle O_{z}\rangle _{y}\langle
O_{x}\rangle _{y},  \label{Correlation}
\end{equation}%
where $O$ is a quantity or property related to each system state \cite%
{vanKampen}, $C_{pf}=\sum_{zx}[P(z,x|y)-P(z|y)P(x|y)]O_{z}O_{x}.$ In here,
indexes $x$ and $z$ run over all possible outcomes occurring at times $t_{x}$
and $t_{z}$ respectively. On the other hand, $y$ index can be any fixed
particular value from all possible outcomes of the second observation.
Markovian processes lead to $C_{pf}=0,$ whatever the conditional state $y$
is. Given that in general $P(z,x|y)=P(z|yx)P(x|y),$ it follows that \textit{%
non-Markovian effects break CPF independence} and are present whenever $%
C_{pf}\neq 0.$ Higher conditional objects are discussed below [Eq.~(\ref%
{HigherCPF})].

\textit{Markovianity of quantum measurements}: The previous memory indicator
can be extended to a quantum regime. In a first step, it is shown that
successive quantum measurement processes fulfill CPF independence. Hence, a
completely isolated quantum system is considered, whose own evolution
between measurements is disregarded. Three consecutive generalized quantum
measurements, which in general are \textit{different and arbitrary}, deliver
the successive random outcomes $x\rightarrow y\rightarrow z.$ The
corresponding measurement operators \cite{milburn} are $x\leftrightarrow
\Omega _{x},$ $y\leftrightarrow \Omega _{y},$ $z\leftrightarrow \Omega _{z}$
(Fig.~1) and satisfy $\sum_{x}\Omega _{x}^{\dagger }\Omega
_{x}=\sum_{y}\Omega _{y}^{\dagger }\Omega _{y}=\sum_{z}\Omega _{z}^{\dagger
}\Omega _{z}=\mathrm{I},$ where $\mathrm{I}$ is the identity matrix and the
sum indexes run over all possible outcomes at each stage.

CPF independence entails the calculation of $P(z,x|y)=P(z|yx)P(x|y)$ [Eq.~(%
\ref{Correlation})]. Given that $x$ is in the past of $y,$ $P(x|y)$ is a 
\textit{retrodicted quantum probability}. Thus, it can be written in terms
of the measurement operator $\Omega _{x}$ and the \textquotedblleft past
quantum state\textquotedblright\ $\Xi \equiv (\rho _{0},E_{y}),$ where $\rho
_{0}$ is the initial density matrix and $E_{y}\equiv \Omega _{y}^{\dagger
}\Omega _{y}$ is the effect operator \cite{molmer,retro}. On the other hand, 
$P(z|yx)$ is a standard \textit{predictive quantum probability}. Hence,%
\begin{equation}
P(z,x|y)=\mathrm{Tr}[\Omega _{z}^{\dagger }\Omega _{z}\rho _{y}]\frac{%
\mathrm{Tr}[E_{y}\Omega _{x}\rho _{0}\Omega _{x}^{\dagger }]}{%
\sum_{x^{\prime }}\mathrm{Tr}[E_{y}\Omega _{x^{\prime }}\rho _{0}\Omega
_{x^{\prime }}^{\dagger }]},  \label{Isolated}
\end{equation}%
where the first and second factors correspond to $P(z|yx)$ and $P(x|y)$
respectively \cite{suple}. Furthermore, $\mathrm{Tr}[\bullet ]$ is the trace
operation, while $\rho _{y}$ is the system state after the $y$-measurement.
When the $y$-measurement is a \textit{projective} one, $\Omega
_{y}=|y\rangle \langle y|,$ being associated to an Hermitian operator $%
\mathbb{O}_{y}=\sum_{y}O_{y}|y\rangle \langle y|,$ it follows $\rho
_{y}=|y\rangle \langle y|.$ This state only depends on the outcome $y,$
while being independent of any former outcome $x.$ Thus, \textit{CPF
independence is fulfilled naturally} [Eq.~(\ref{Indenpendence})].
Introducing a \textquotedblleft causal break\textquotedblright\ \cite{modi}
or \textquotedblleft preparation\textquotedblright\ \cite{preparation}, this
property is also valid for non-projective $y$-measurements $[\Omega
_{y}^{\dagger }\Omega _{y}\neq \Omega _{y}]$ \cite{suple}.

\begin{figure}[tbp]
\includegraphics[bb=25 77 735
307,angle=0,width=8.5cm]{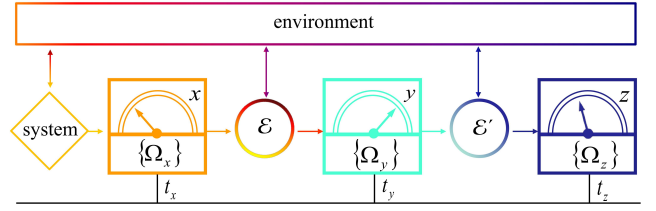}
\caption{Measurement scheme. At times $t_{x}<t_{y}<t_{z}$ an open system is
subjected to three measurement processes whose random outcomes are $%
x\rightarrow y\rightarrow z.$ A set of operators $\{\Omega _{x}\},$ $%
\{\Omega _{y}\},$ and $\{\Omega _{z}\}$ define the measurement processes in
a quantum regime. $\mathcal{E}$ and $\mathcal{E}^{\prime }$ are the system
propagators between consecutive measurements.}
\end{figure}

\textit{Quantum Markovian dynamics}: In general, the system evolves between
consecutive measurement events. Its dynamics is defined as Markovian if, for
arbitrary measurement processes, it does not break CPF independence. This
condition is preserved when the system propagator does not depend on past
measurement outcomes. Propagator independence of future outcomes is
guaranteed by causality. Hence, from Eq.~(\ref{Isolated}) the CPF
probability reads%
\begin{equation}
P(z,x|y)\!=\!\mathrm{Tr}(\Omega _{z}^{\dagger }\Omega _{z}\mathcal{E}%
^{\prime }[\rho _{y}])\frac{\mathrm{Tr}(E_{y}\mathcal{E}[\Omega _{x}\rho
_{0}\Omega _{x}^{\dagger }])}{\sum_{x^{\prime }}\mathrm{Tr}(E_{y}\mathcal{E}%
[\Omega _{x^{\prime }}\rho _{0}\Omega _{x^{\prime }}^{\dagger }])},
\label{QuantumMarkov}
\end{equation}%
where $\mathcal{E}=\mathcal{E}(t_{y},t_{x})$\ and $\mathcal{E}^{\prime }=%
\mathcal{E}^{\prime }(t_{z},t_{y})$\ are the\ (measurement independent)
system density matrix propagators between consecutive events (Fig.~1). The
fulfillment of condition (\ref{QuantumMarkov}) provides an explicit
measurement-based criteria for defining quantum Markovianity, which
similarly to classical systems, leads to a vanishing CPF correlation (\ref%
{Correlation}). In particular, a unitary dynamics is Markovian.

\textit{Quantum system-environment models}: Consider a system $(s)$
interacting with its environment $(e),$ with total Hamiltonian $H_{T}.$ The
dynamics is sets by the propagator%
\begin{equation}
\mathcal{E}_{t}=\exp (t\mathcal{L}_{se}),\ \ \ \ \ \ \mathcal{L}%
_{se}[\bullet ]=-i[H_{T},\bullet ].  \label{Exponential}
\end{equation}%
The system density matrix is $\rho _{t}=\mathrm{Tr}_{e}(\mathcal{E}_{t}[\rho
_{0}^{se}]),$ where $\rho _{0}^{se}$ is the initial system-environment
state. For measurements that \textit{only provide information about system
observables}, the proposed scheme (Fig.~1)\ allows us to characterizing
departures of the system partial dynamics from a Markovian regime. The
probabilities calculus is almost the same as for Eq.~(\ref{QuantumMarkov}) 
\cite{suple}. In particular, after the second measurement $(y)$ the
bipartite state $\rho ^{se}$\ suffers the disruptive transformation $\rho
^{se}\rightarrow \rho _{y}\otimes \sigma _{e}^{yx}.$ Thus, the system and
the environment become uncorrelated. This property is always granted by
projective measurements. The system state $\rho _{y}$ does not depend on the
past measurement outcome $x,$ while the marginal bath state $\sigma
_{e}^{yx} $ does. It is given by 
\begin{equation}
\sigma _{e}^{yx}=\frac{\mathrm{Tr}_{s}(E_{y}\mathcal{E}_{t}[\Omega _{x}\rho
_{0}^{se}\Omega _{x}^{\dagger }])}{\mathrm{Tr}_{se}(E_{y}\mathcal{E}%
_{t}[\Omega _{x}\rho _{0}^{se}\Omega _{x}^{\dagger }])}.  \label{BathState}
\end{equation}%
The CPF probability, similarly to Eq. (\ref{QuantumMarkov}), is \cite{suple} 
\begin{eqnarray}
&&P(z,x|y)=\mathrm{Tr}_{se}(\Omega _{z}^{\dagger }\Omega _{z}\mathcal{E}%
_{\tau }[\rho _{y}\otimes \sigma _{e}^{yx}])  \notag \\
&&\ \ \ \ \ \ \ \ \ \ \ \ \ \ \times \frac{\mathrm{Tr}_{se}(E_{y}\mathcal{E}%
_{t}[\Omega _{x}\rho _{0}^{se}\Omega _{x}^{\dagger }])}{\sum_{x^{\prime }}%
\mathrm{Tr}_{se}(E_{y}\mathcal{E}_{t}[\Omega _{x^{\prime }}\rho
_{0}^{se}\Omega _{x^{\prime }}^{\dagger }])},  \label{PCondBipartita}
\end{eqnarray}%
where $t\equiv t_{y}-t_{x}$ and $\tau \equiv t_{z}-t_{y}$ are the time
intervals between consecutive measurements. Given the dependence of the
environment state $\sigma _{e}^{yx}$ on the first measurement $(x),$ here
the CPF independence is broken in general. The properties of this departure
can be quantified with the CPF correlation (\ref{Correlation}), $%
C_{pf}\rightarrow C_{pf}(t,\tau ),$ which can be obtained from the previous
expression and the\ system observables definition.

\textit{Born-Markov approximation}: A Markovian regime, defined by the
measurement-based condition (\ref{QuantumMarkov}), is approached when the
initial bipartite state is separable, $\rho _{0}^{se}=\rho _{0}\otimes
\sigma _{e},$ and for arbitrary time $t,$%
\begin{equation}
\mathcal{E}_{t}[\Omega _{x}\rho _{0}^{se}\Omega _{x}^{\dagger }]\simeq 
\tilde{\rho}_{x}(t)\otimes \sigma _{e},  \label{BornMarkov}
\end{equation}%
where $\tilde{\rho}_{x}(0)=\Omega _{x}\rho _{0}\Omega _{x}^{\dagger }.$
Indeed, under this approximation the bath state is (approximately)
unperturbed during the total evolution, $\sigma _{e}^{yx}\simeq \sigma _{e}$
[see Eq. (\ref{BathState})], implying $C_{pf}(t,\tau )\simeq 0.$ Therefore, 
\textit{the CPF correlation measures and quantifies departures with respect
to the standard Born-Markov approximation.} In fact, the separability
constraint (\ref{BornMarkov}) is valid when the conditions under which it
applies are fulfilled \cite{cohen}.

\textit{Classical environment fluctuations}: Instead of a unitary bipartite
evolution [Eq.~(\ref{Exponential})], the open system dynamics may be
described by a quantum Liouville operator $\mathcal{L}_{st}(t)$ modulated by
classical noise fluctuations,%
\begin{equation}
\frac{d}{dt}\rho _{t}^{st}=-i\mathcal{L}_{st}(t)[\rho _{t}^{st}].
\label{LiouvilleSt}
\end{equation}%
The system density operator $\rho _{t}=\overline{\rho _{t}^{st}}$ follows by
averaging over realizations of $\mathcal{L}_{st}(t)$ [overbar symbol]. The
CPF probability can straightforwardly be written as%
\begin{equation}
P(z,x|y)\!=\!\overline{P_{st}(z,x}|y),  \label{CPFStochastic}
\end{equation}%
where the classical average is restricted to the $y$-outcome and the
\textquotedblleft stochastic probability\textquotedblright\ $P_{st}(z,x|y)$
follows from Eq.~(\ref{QuantumMarkov}) under the replacements $\mathcal{E}%
\rightarrow \exp [-i\int_{0}^{t}dt^{\prime }\mathcal{L}_{st}(t^{\prime })]$
and $\mathcal{E}^{\prime }\rightarrow \exp [-i\int_{t}^{t+\tau }dt^{\prime }%
\mathcal{L}_{st}(t^{\prime })].$ \textit{Non-Markovian effects are then
related to the correlation between both intermediate propagators, while
white fluctuations lead to a Markovian dynamics} \cite{suple}. The model (%
\ref{LiouvilleSt}) not only covers the case of stochastic Hamiltonian
evolutions \cite{GaussianNoise} but also quantum-classical hybrid
arrangements \cite{molmer,wiseman} where in general the incoherent and
quantum systems may affect each other \cite{LindbladRate}.

\textit{CPF correlation properties}: Similarly to classical systems, a
non-Markovian regime is defined by the condition $C_{pf}(t,\tau )\gtrless 0.$
In general $C_{pf}(t,\tau )\neq C_{pf}(\tau ,t).$ From Eq. (\ref{BathState})
[and (\ref{CPFStochastic})] it follows $\lim_{\tau \rightarrow
0}C_{pf}(t,\tau )=0$ and $\lim_{t\rightarrow 0}C_{pf}(t,\tau )=0,$ this last
condition\ being only valid when the system and the environment are
uncorrelated at the initial time. If the environment fluctuations have a 
\textit{finite correlation time} $\tau _{c}$ \cite{breuerbook}, $%
C_{pf}(t,\tau )\simeq ~0$ if $t\gg \tau _{c}$ or $\tau \gg \tau _{c}.$ Thus, 
$\lim_{\tau \rightarrow \infty }C_{pf}(t,\tau )=\lim_{t\rightarrow \infty
}C_{pf}(t,\tau )=\lim_{t\rightarrow \infty }C_{pf}(t,ct)=0,$ $\forall c>0.$
In an experimental setup $C_{pf}(t,\tau )$ follows straightforwardly by
performing a statistical average with a \textit{post-selected} sub-ensemble
of realizations $x\rightarrow y^{\prime }\rightarrow z,$ where $y^{\prime }$
is the chosen conditional fixed value. Contrarily to classical systems, the
condition $C_{pf}(t,\tau )\neq 0$ \textit{may depends on the chosen
measurement observables.} This reacher behavior in turn gives a deeper
characterization of memory effects in quantum systems.

\textit{Higher order CPF correlations}: The CPF correlation (\ref%
{Correlation})\ can be generalized by increasing the number of observations, 
$x\rightarrow y_{1}\rightarrow y_{2}\rightarrow \cdots y_{n}\rightarrow z.$
An $n$-order CPF correlation is defined as%
\begin{equation}
C_{pf}^{(n)}=\sum_{zx}[P(z,x|\mathbf{y})-P(z|\mathbf{y})P(x|\mathbf{y}%
)]O_{z}O_{x},  \label{HigherCPF}
\end{equation}%
where $\mathbf{y\equiv }y_{n},\cdots ,y_{2},y_{1}.$ Given that $%
P(z,x|y_{n}\cdots y_{1})=P(z|y_{n}\cdots y_{1},x)P(x|y_{n}\cdots y_{1}),$ $%
C_{pf}^{(n)}$ is sensitive to memory effects that may only appear in these
higher conditional objects. For example, it may happen that $C_{pf}^{(k)}=0$ 
$\forall k<n$ and $C_{pf}^{(n)}\neq 0.$ Classical Markovian processes
fulfill $C_{pf}^{(n)}=0$ $\forall n.$ Thus, higher order CPF correlations
provide an overall check of the definition of classical Markovianity in
terms of conditional probabilities. This property guarantees the consistence
of the present formalism with Ref. \cite{modi}. In fact, $C_{pf}^{(n)}$ can
also be extended and calculated in a quantum regime \cite{suple} (previous
expressions correspond to $n=1).$ Nevertheless, in contrast to classical
systems, given the degrees of freedom provided by the measurement operators,
for a wide class of quantum dynamics [Eqs.~(\ref{Exponential}) and (\ref%
{LiouvilleSt})] it is expected that $C_{pf}^{(1)}\neq 0$ \cite{suple}. Thus,
memory effects can be analyzed over the basis of a minimal three
quantum-measurements scheme (Fig.~1). The next results support this
conclusion.

\textit{Dephasing spin bath}: As a first example we consider a paradigmatic
model of decoherence \cite{Zurek,ZurekRMP,Paz}: a qubit system interacting
with a $N$-spin bath via the microscopic interaction Hamiltonian%
\begin{equation}
H_{T}=\sigma _{\hat{z}}\otimes \sum_{k=1}^{N}g_{k}\sigma _{\hat{z}}^{(k)}.
\label{SpinBath}
\end{equation}%
Here $\sigma _{\hat{z}}$ is the system Pauli matrix in the $\hat{z}$%
-direction (Bloch sphere) \cite{comment1}, whose eigenvectors are denoted as 
$|\pm \rangle .$ On the other hand, $\sigma _{\hat{z}}^{(k)}$ is $\hat{z}-$%
Pauli matrix corresponding to the $k$-spin. Its eigenvectors are denoted by $%
|\uparrow \rangle _{k}$ and $|\downarrow \rangle _{k}.$ $\{g_{k}\}$ are real
coupling constants.

As is well known, the model (\ref{SpinBath}) admits an exact solution \cite%
{Zurek,ZurekRMP,Paz}. Assuming a separable pure initial condition $\rho
_{0}^{se}=|\Psi _{0}^{se}\rangle \langle \Psi _{0}^{se}|,$ where%
\begin{equation}
|\Psi _{0}^{se}\rangle =(a|+\rangle +b|-\rangle )\otimes
\sum_{k=1}^{N}(\alpha _{k}|\uparrow \rangle _{k}+\beta _{k}|\downarrow
\rangle _{k}),  \label{CI}
\end{equation}%
the system density matrix reads $\rho _{t}=|a|^{2}|+\rangle \langle
+|+|b|^{2}|-\rangle \langle -|+ab^{\ast }c_{t}|+\rangle \langle -|+a^{\ast
}bc_{t}^{\ast }|-\rangle \langle +|.$ Its evolution can then be written as%
\begin{equation}
\frac{d\rho _{t}}{dt}=\frac{-i}{2}\omega (t)[\sigma _{\hat{z}},\rho
_{t}]+\gamma (t)\frac{1}{2}(\sigma _{\hat{z}}\rho _{t}\sigma _{\hat{z}}-\rho
_{t}),  \label{Lindblad}
\end{equation}%
where the time dependent frequency $\omega (t)$ and decay rate $\gamma (t)$
follow from $\gamma (t)+i\omega (t)=-(1/c_{t})(d/dt)c_{t}.$ The system
coherence behavior,%
\begin{equation}
c_{t}=\prod_{k=1}^{N}(|\alpha _{k}|^{2}e^{+i2g_{k}t}+|\beta
_{k}|^{2}e^{-i2g_{k}t}),  \label{CoherenceDecay}
\end{equation}%
depends on the initial bath state and coupling constants.

\textit{Measurement scheme and CPF correlation}: In order to check
non-Markovian effects, the three measurements (Fig.~1) are chosen as
projective ones, being performed in $\hat{x}$-direction. The outcomes of
each measurement are then $x=\pm 1,$ $y=\pm 1,$ $z=\pm 1,$ which in turn
define the system operators values in Eq.~(\ref{Correlation}), $O_{z}=z$ and 
$O_{x}=x.$ The measurement operators are $\{\Omega _{x}\}=\{\Omega
_{y}\}=\{\Omega _{z}\}=|\hat{x}_{\pm }\rangle \langle \hat{x}_{\pm }|,$
where $|\hat{x}_{\pm }\rangle =(|+\rangle \pm |-\rangle )/\sqrt{2}.$

All calculations leading to the CPF probability (\ref{PCondBipartita}) can
be performed in an exact way \cite{Escort}. Assuming, for simplicity, that
the system begin in the state $|+\rangle $ $(a=1,b=0),$%
\begin{equation}
P(zx|y)=\frac{1}{4}[1+xyf(t)+zyf(\tau )+zxf(t,\tau )],
\label{CPFProbability}
\end{equation}%
where $f(t)=\mathrm{Re}[c_{t}]$ gives the coherence decay and $f(t,\tau
)=[f(t+\tau )+f(t-\tau )]/2.$ From here, it follows $\langle O_{z}\rangle
_{y}=yf(\tau ),$ $\langle O_{x}\rangle _{y}=yf(t),$ and $\langle
O_{z}O_{x}\rangle _{y}=f(t,\tau ).$ The exact expression for the CPF
correlation (\ref{Correlation}) then is%
\begin{equation}
C_{pf}(t,\tau )=f(t,\tau )-f(t)f(\tau ),  \label{CPFExactSpin}
\end{equation}%
which, due to the symmetries, here is independent of the conditional value $%
y=\pm 1.$

\textit{A non-Markovian quantum dynamical semigroup}: As is well known \cite%
{ZurekRMP,Paz}, the model (\ref{SpinBath}) may leads to Gaussian system
decay behaviors. For example, taking $g_{k}=\frac{1}{\sqrt{N}}g,$ $\alpha
_{k}=\beta _{k}=1/2,$ for $N\gg 1$ it follows $c_{t}\simeq \exp [-2(gt)^{2}]$
(behavior valid before the unitary recurrence time). Thus, $\omega (t)=0$
and $\gamma (t)\simeq 4g^{2}t.$ This \textit{positive time-dependent rate }%
leads to a \textit{time-dependent Lindblad semigroup} [Eq. (\ref{Lindblad})]
that, in almost all proposed non-Markovian measure schemes \cite%
{BreuerReview,plenioReview}, is classified as a Markovian evolution. In
contrast, here due to strong departures from condition (\ref{BornMarkov}),
the CPF correlation does not vanish. In fact, for $N\gg 1,$ it can be
approximated~as%
\begin{equation}
C_{pf}(t,\tau )\!\simeq \!\frac{e^{-2g^{2}(t+\tau )^{2}}+e^{-2g^{2}(t-\tau
)^{2}}}{2}-e^{-2g^{2}(t^{2}+\tau ^{2})}.  \label{CPFInfinite}
\end{equation}%
In Fig. 2 (left panels) we plot $C_{pf}(t,\tau ),$ which is very well fitted
by the previous expression. The symmetry $C_{pf}(t,\tau )=C_{pf}(\tau ,t)$
is a consequence of the chosen environment initial conditions. Furthermore,
for increasing equal time intervals $C_{pf}(t,t)\simeq 1/2.$ This property
indicates that the bath correlation does not decay (vanishes) in time
(infinite bath correlation time). 
\begin{figure}[tbp]
\includegraphics[bb=0 0 302 310,angle=0,width=3.5cm]{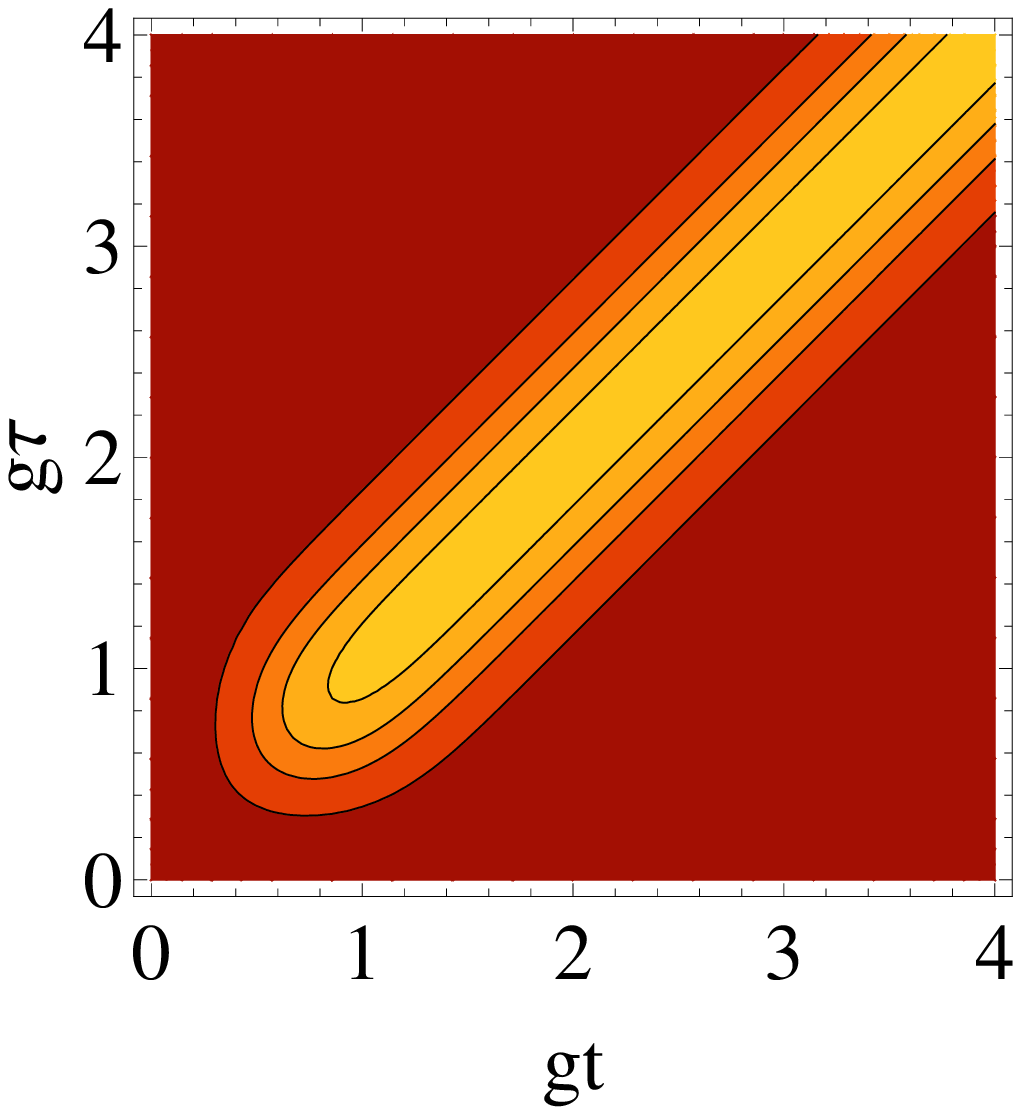} %
\includegraphics[bb=0 0 302 310,angle=0,width=3.5cm]{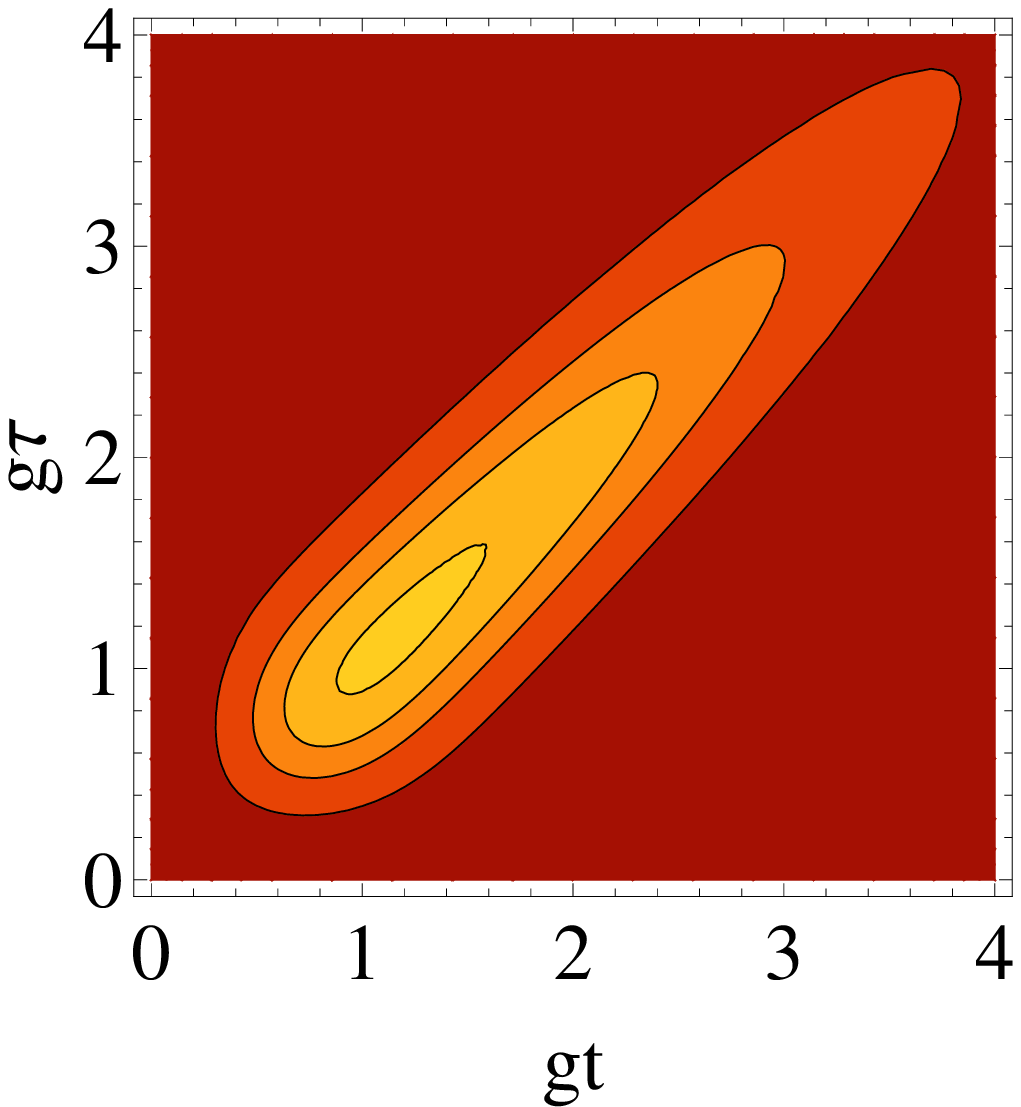} %
\includegraphics[bb=0 0 320 255,angle=0,width=3.5cm]{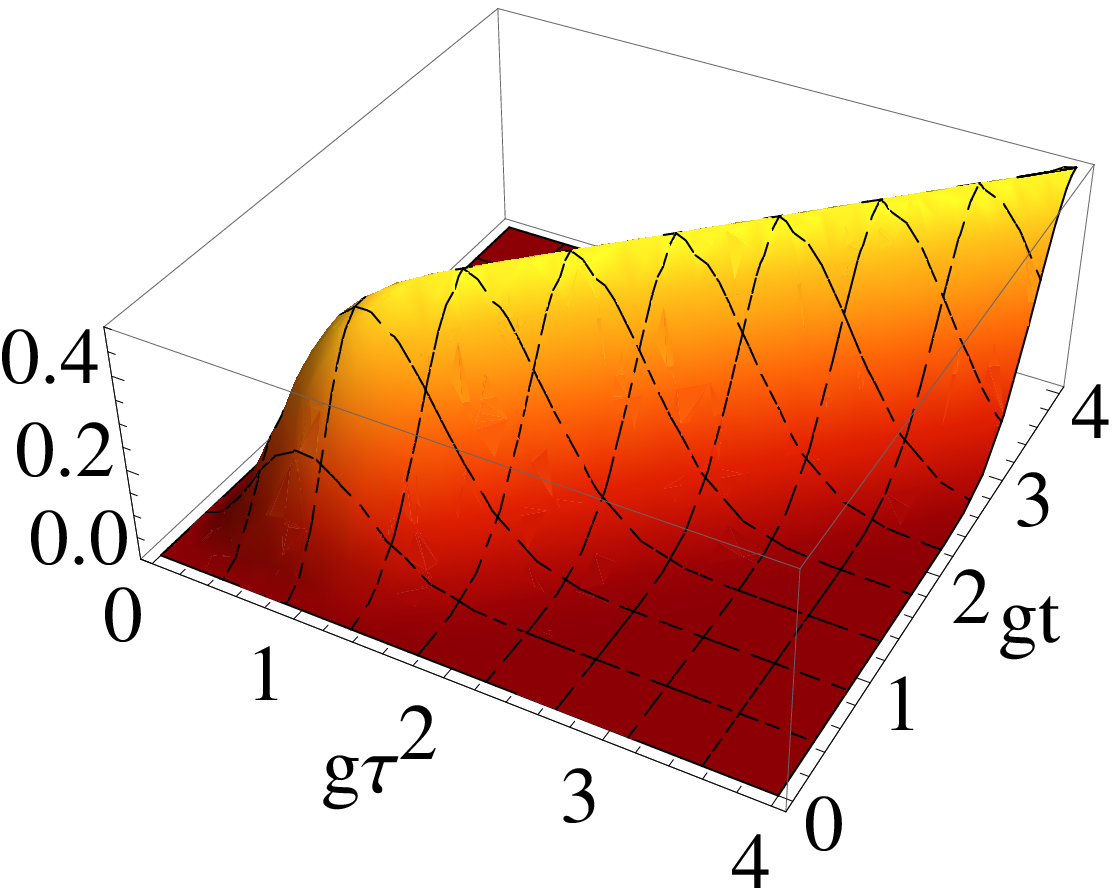} %
\includegraphics[bb=0 0 320 255,angle=0,width=3.5cm]{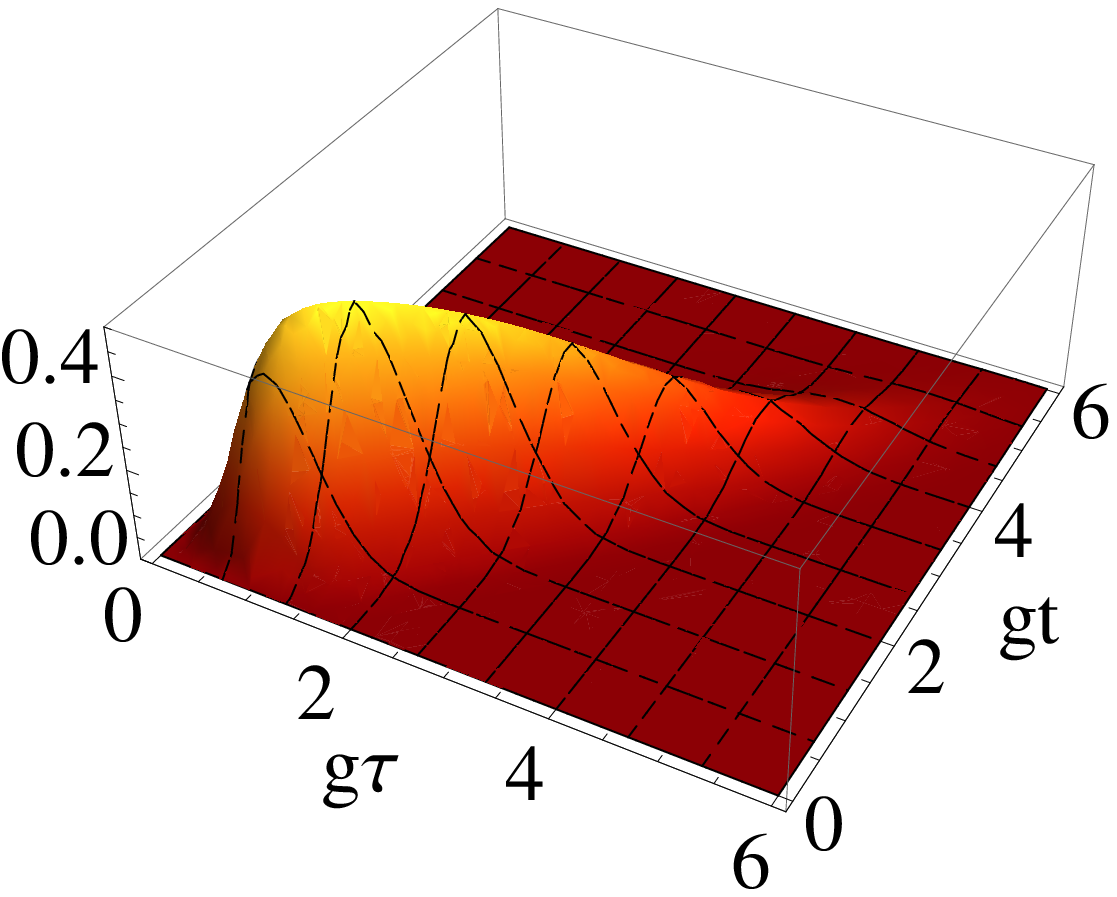}
\caption{Left panels: CPF correlation (\protect\ref{CPFExactSpin}) for the
spin bath model (\protect\ref{SpinBath}), with coupling $g_{k}=g/\protect%
\sqrt{N}.$ The parameters of the initial condition (\protect\ref{CI}) are $%
a=1,$ $b=0,$ $\protect\alpha _{k}=\protect\beta _{k}=1/2,$ and $N=50.$ Right
panels: CPF correlation for the stochastic Hamiltonian model (\protect\ref%
{Hst}), with noise correlation $\overline{\protect\xi _{t}\protect\xi %
_{t^{\prime }}}=g^{2}\exp [-|t-t^{\prime }|/\protect\tau _{c}].$ The system
starts at the same initial condition. The parameters are $\protect\tau %
_{c}g=100.$ For $\protect\tau _{c}\rightarrow \infty ,$ the left panels are
recovered.}
\end{figure}

\textit{Stochastic Hamiltonian}: An alternative decoherence model, which
mimics the interaction with a spin bath \cite{anderson}, is given by a
stochastic Hamiltonian evolution%
\begin{equation}
\mathcal{L}_{st}(t)[\bullet ]=-i\xi _{t}[\sigma _{\hat{z}},\bullet ],
\label{Hst}
\end{equation}%
where $\xi _{t}$\ is a classical noise [Eq.~(\ref{LiouvilleSt})]. The
density matrix evolution is also defined by Eq. (\ref{Lindblad}), where now%
\begin{equation}
c_{t}=\overline{\exp \Big{[}-2i\int_{0}^{t}dt^{\prime }\xi (t^{\prime })%
\Big{]}}.  \label{c(t)Average}
\end{equation}%
The CPF probability (\ref{CPFStochastic}) can also be calculated in an exact
way \cite{Escort}. It can be written as in Eq. (\ref{CPFProbability}) where
similarly $f(t)=\mathrm{Re}[c_{t}]$ [Eq. (\ref{c(t)Average})] while $%
f(t,\tau )=\overline{\mathrm{Re}[\exp [-2i\int_{0}^{t}dt^{\prime }\xi
(t^{\prime })]\mathrm{Re}[\exp [-2i\int_{t}^{t+\tau }dt^{\prime }\xi
(t^{\prime })]}.$

Taking a Gaussian noise with $\overline{\xi _{t}}=0$ and correlation $%
\overline{\xi _{t}\xi _{t^{\prime }}}=g^{2}\exp [-|t-t^{\prime }|/\tau
_{c}], $ Eq.~(\ref{Lindblad}) is defined with $\omega (t)=0$ and $\gamma
(t)=4g^{2}\tau _{c}(1-e^{-t/\tau _{c}})>0,$ providing a second example of a
non-Markovian time-dependent quantum semigroup. In particular, in the limit $%
\tau _{c}\rightarrow \infty ,$ the same Gaussian behavior is recovered, $%
c_{t}=\exp [-2(gt)^{2}].$ Thus, the CPF correlation is exactly given by Eq. (%
\ref{CPFInfinite}) [left panels in Fig. 2]. On the other hand, taking $%
\gamma _{w}/2=g^{2}\tau _{c}$ as a constant parameter, in the limit $\tau
_{c}\rightarrow 0,$ a Markovian regime is achieved, $C_{pf}(t,\tau
)\rightarrow 0,$ with $c_{t}=\exp [-2\gamma _{w}t].$ In Fig. 2 (right
panels), we also plot $C_{pf}(t,\tau )$ for a finite $\tau _{c}.$ All
expected characteristics corresponding to a finite bath correlation time are
developed.

\textit{Conclusions}:\ Similarly to classical systems, a quantum
(memoryless) Markovian regime was defined by the statistical independence of
past and future events when conditioned to a present system state. Thus, a
minimal set of three time-ordered quantum measurements leads to an
operational (measurement-based) definition of quantum non-Markovianity.
Post-selection gives the conditional character of the measurement scheme.
Its associated CPF correlation is a direct and univocal indicator of
departures from Born-Markov and white noise approximations.

The proposed scheme leads to a powerful theoretical and experimental basis
for the study of memory effects in open quantum systems. Its capacity for
characterizing the underlying physical origin of memory effects was
established by studying different dephasing mechanisms that admit an exact
treatment. The conditional character of the measurement scheme opens an
interesting way to describe quantum memory effects by means of\ recent
theoretical and experimental advances in retrodicted quantum measurement
processes \cite%
{vaidman,murch,molmer,haroche,huard,naghi,decay,retro,wiseman,dressel,barnett}.

\textit{Acknowledgments}: The author thanks to M. M. Guraya for a critical
reading of the manuscript. This paper was supported by Consejo Nacional de
Investigaciones Cient\'{\i}ficas y T\'{e}cnicas (CONICET), Argentina.

\begin{widetext}
\begin{center}
\Huge{Supplemental Material}
\end{center}

\appendix

\section{Markovianity of quantum measurements}

The classical notion of Markovianity \cite{vanKampen}\ can be defined
through a conditional past-future (CPF) statistical independence \cite%
{CoverTomas}. Given a set of three arbitrary time-ordered events $%
x\rightarrow y\rightarrow z,$ their joint probability $P(z,y,x)$ can
alternatively be written as $P(z,y,x)=P(z|y,x)P(y,x),$ and $%
P(z,y,x)=P(z,x|y)P(y),$ where in general $P(b|a)$ denotes the conditional
probability of $b$ given $a.$ Using that $P(y,x)=P(x|y)P(y),$ it follows%
\begin{equation}
P(z,x|y)=P(z|y,x)P(x|y).  \label{CPFProba}
\end{equation}%
For Markovian processes, $P(z|y,x)\rightarrow P(z|y),$ which recover their
definition in terms of a CPF statistical independence \cite{CoverTomas}, $%
P(z,x|y)=P(z|y)P(x|y).$

Here, the successive events $x\rightarrow y\rightarrow z$ correspond to the
outcomes of three consecutive generalized quantum measurements \cite{milburn}%
, defined in terms of the measurement operators $\{\Omega _{x}\},$ $\{\Omega
_{y}\},$ and $\{\Omega _{z}\}.$ The calculus of the previous probabilities
is as follows. At each measurement step the system state suffers the
transformation \cite{milburn}%
\begin{equation}
\rho _{i}\rightarrow \rho _{j}=\frac{\Omega _{j}\rho _{i}\Omega
_{j}^{\dagger }}{\mathrm{Tr}[\Omega _{j}^{\dagger }\Omega _{j}\rho _{i}]},\
\ \ \ \ \ \ P(j|i)=\mathrm{Tr}[\Omega _{j}^{\dagger }\Omega _{j}\rho _{i}],
\end{equation}%
where $i=0,x,y$ and respectively $j=x,y,z.$ Furthermore, $\mathrm{Tr}%
[\bullet ]$ is the trace operation, while $p(j|i)$ is the conditional
probability of outcome $j$ given that the previous state is $\rho _{i},$
associated to outcome $i,$ while $\rho _{0}$ is the initial state.

After the first $x$-measurement, it occurs the transformation $\rho
_{0}\rightarrow \rho _{x},$ where%
\begin{equation}
\rho _{x}=\frac{\Omega _{x}\rho _{0}\Omega _{x}^{\dagger }}{\mathrm{Tr}%
[\Omega _{x}^{\dagger }\Omega _{x}\rho _{0}]},\ \ \ \ \ \ P(x|0)=\mathrm{Tr}%
[\Omega _{x}^{\dagger }\Omega _{x}\rho _{0}].
\end{equation}%
Here $P(x|0)$ is the probability of outcomes $x$ given the initial state $%
\rho _{0}.$ After the second $y$-measurement, the transformation $\rho
_{x}\rightarrow \rho _{y}$ occurs, where now%
\begin{equation}
\rho _{y}=\frac{\Omega _{y}\rho _{x}\Omega _{y}^{\dagger }}{\mathrm{Tr}%
[\Omega _{y}^{\dagger }\Omega _{y}\rho _{x}]},\ \ \ \ P(y|x)=\mathrm{Tr}%
[\Omega _{y}^{\dagger }\Omega _{y}\rho _{x}].
\end{equation}%
The joint probability of both measurement outcomes $P(y,x)=P(y|x)P(x|0),$
can be written as%
\begin{equation}
P(y,x)=\mathrm{Tr}[\Omega _{y}^{\dagger }\Omega _{y}\Omega _{x}\rho
_{0}\Omega _{x}^{\dagger }].
\end{equation}%
In consequence, the quantum retrodicted probability $P(x|y)=$ $P(y,x)/P(y),$
using that $P(y)=\sum_{x}P(y,x),$ reads%
\begin{equation}
P(x|y)=\frac{\mathrm{Tr}[\Omega _{y}^{\dagger }\Omega _{y}\Omega _{x}\rho
_{0}\Omega _{x}^{\dagger }]}{\sum_{x^{\prime }}\mathrm{Tr}[\Omega
_{y}^{\dagger }\Omega _{y}\Omega _{x^{\prime }}\rho _{0}\Omega _{x^{\prime
}}^{\dagger }]},  \label{PXYAp}
\end{equation}%
where the index $x^{\prime }$ runs over all possible outcomes of the first $%
x $-measurement. This\ expression recovers the result of a past quantum
state formalism \cite{molmer,retro}.

For projective $y$-measurements, the state $\rho _{y}$ does not depend on
the previous outcome $x.$ When the $y$-measurement is a general one $[\Omega
_{y}^{\dagger }\Omega _{y}\neq \Omega _{y}],$ $\rho _{y}$ also depends on
the previous outcome $x$ producing an \textit{artificial}
measurement-induced violation of CPF independence. This effect can always be
surpassed as follows. An extra projective measurement is performed (after
the second one), being defined by a set of projectors $\{\Pi _{\alpha
}=|\alpha \rangle \langle \alpha |\},$ which satisfy $\Pi _{\alpha }\Pi
_{\alpha ^{\prime }}=\delta _{\alpha \alpha ^{\prime }}\Pi _{\alpha },$ and $%
\sum_{\alpha }\Pi _{\alpha }=I.$ Thus, given the outcome $\alpha ,$ $\rho
_{y}\rightarrow \Pi _{\alpha }=|\alpha \rangle \langle \alpha |.$ Depending
of the original $y$-outcome an extra \textit{conditional unitary rotation} $%
\mathcal{R}(y|\alpha )$ is applied such that for all $\alpha ,$%
\begin{equation}
\rho _{y}\rightarrow \rho _{y}=\mathcal{R}(y|\alpha )[\Pi _{\alpha
}]=|y\rangle \langle y|,  \label{Prepa}
\end{equation}%
where the states $\{|y\rangle \}$ are arbitrary ones. In this way each
outcome $y$ has assigned a past-independent state. Adding an extra random
degree of freedom, the refreshed state $\rho _{y}$ may also correspond to a
mixed state. The role of this \textquotedblleft causal
break\textquotedblright\ \cite{modi} or \textquotedblleft
preparation\textquotedblright\ \cite{preparation} is to \textquotedblleft
erase the dependence of the system evolution on the previous history of the
system without erasing the memory of the universe, i.e., the
system-environment arrangement,\textquotedblright\ where here the
environment is defined by the bipartite intrinsic nature of the intermediate
generalized measurement \cite{milburn}.

After the final $z$-measurement, $\rho _{y}\rightarrow \rho _{z},$ where%
\begin{equation}
\rho _{z}=\frac{\Omega _{z}\rho _{y}\Omega _{z}^{\dagger }}{\mathrm{Tr}%
[\Omega _{z}^{\dagger }\Omega _{z}\rho _{y}]},\ \ \ \ \ P(z|y,x)=\mathrm{Tr}%
[\Omega _{z}^{\dagger }\Omega _{z}\rho _{y}].  \label{PZYXAp}
\end{equation}%
Given that $P(z|y,x)=P(z|y),$ the Markovian CPF independence is fulfilled, $%
P(z,x|y)=P(z|y)P(x|y),$ with%
\begin{equation}
P(z,x|y)=\mathrm{Tr}[\Omega _{z}^{\dagger }\Omega _{z}\rho _{y}]\frac{%
\mathrm{Tr}[E_{y}\Omega _{x}\rho _{0}\Omega _{x}^{\dagger }]}{%
\sum_{x^{\prime }}\mathrm{Tr}[E_{y}\Omega _{x^{\prime }}\rho _{0}\Omega
_{x^{\prime }}^{\dagger }]},  \label{CPFProbaAp}
\end{equation}%
where $E_{y}\equiv \Omega _{y}^{\dagger }\Omega _{y}.$ When the intermediate 
$y$-measurement is a projective one, the extra quantum operations defined by
Eq. (\ref{Prepa}) are unnecessary.

\section{Quantum system-environment models}

For a bipartite system-environment arrangement the CPF probability can be
calculated as follows. In all steps, it is assumed that the measurement
processes are performed only on the system, $\{\Omega _{j}\}\leftrightarrow
\{\Omega _{j}\otimes \mathrm{I}_{e}\},$ where $j=x,y,z$ and $\mathrm{I}_{e}$
is the identity matrix in the environment Hilbert space.

Denoting with $\rho _{0}^{se}$ the initial bipartite state, after the first $%
x$-measurement it occurs the transformation $\rho _{0}^{se}\rightarrow \rho
_{x}^{se},$ where%
\begin{equation}
\rho _{x}^{se}=\frac{\Omega _{x}\rho _{0}^{se}\Omega _{x}^{\dagger }}{%
\mathrm{Tr}_{se}(\Omega _{x}^{\dagger }\Omega _{x}\rho _{0}^{se})}.
\end{equation}%
The probability of each outcome is%
\begin{equation}
P(x|0)=\mathrm{Tr}_{se}(\Omega _{x}^{\dagger }\Omega _{x}\rho _{0}^{se}).
\end{equation}%
During a time interval$\ t=t_{y}-t_{x},$ the bipartite arrangement evolves
with the propagator $\mathcal{E}_{t}=\exp (t\mathcal{L}_{se}),$ where $%
\mathcal{L}_{se}[\bullet ]=-i[H_{T},\bullet ]$ induce a completely positive
bipartite dynamics. After the second $y$-measurement, it follows the
transformation $\mathcal{E}_{t}[\rho _{x}^{se}]\rightarrow \rho _{y}^{se},$
where%
\begin{equation}
\rho _{y}^{se}=\frac{\Omega _{y}\mathcal{E}_{t}[\rho _{x}^{se}]\Omega
_{y}^{\dagger }}{\mathrm{Tr}_{se}(\Omega _{y}^{\dagger }\Omega _{y}\mathcal{E%
}_{t}[\rho _{x}^{se}])}.  \label{RhoYBipartito}
\end{equation}%
The conditional probability of outcome $y$ given that the previous one was $%
x $ is%
\begin{equation}
P(y|x)=\mathrm{Tr}_{se}(\Omega _{y}^{\dagger }\Omega _{y}\mathcal{E}%
_{t}[\rho _{x}^{se}]).
\end{equation}%
Thus, the joint probability for both measurement outcomes, $%
P(y,x)=P(y|x)P(x|0),$ is 
\begin{equation}
P(y,x)=\mathrm{Tr}_{se}(\Omega _{y}^{\dagger }\Omega _{y}\mathcal{E}%
_{t}[\Omega _{x}\rho _{0}^{se}\Omega _{x}^{\dagger }]).
\end{equation}%
Using Bayes rule, the retrodicted probability $P(x|y)=$ $P(y,x)/P(y),$ here
reads%
\begin{equation}
P(x|y)=\frac{\mathrm{Tr}_{se}(\Omega _{y}^{\dagger }\Omega _{y}\mathcal{E}%
_{t}[\Omega _{x}\rho _{0}^{se}\Omega _{x}^{\dagger }])}{\sum_{x^{\prime }}%
\mathrm{Tr}_{se}(\Omega _{y}^{\dagger }\Omega _{y}\mathcal{E}_{t}[\Omega
_{x^{\prime }}\rho _{0}^{se}\Omega _{x^{\prime }}^{\dagger }])}.
\label{RetroBipartita}
\end{equation}

At this stage, for projective $y$-measurements $[\Omega _{y}=|y\rangle
\langle y|]$ the state (\ref{RhoYBipartito}) is a separable one, $\rho
_{y}^{se}=\rho _{y}\otimes \sigma _{e}^{yx},$ where the bath state can be
written as%
\begin{equation}
\sigma _{e}^{yx}=\mathrm{Tr}_{s}(\rho _{y}^{se})=\frac{\mathrm{Tr}%
_{s}(\Omega _{y}^{\dagger }\Omega _{y}\mathcal{E}_{t}[\rho _{x}^{se}])}{%
\mathrm{Tr}_{se}(\Omega _{y}^{\dagger }\Omega _{y}\mathcal{E}_{t}[\rho
_{x}^{se}])}.  \label{BathStateYX}
\end{equation}%
For generalized measurements $[\Omega _{y}^{\dagger }\Omega _{y}\neq \Omega
_{y}]$ this feature can also be induced with the previous extra operations
[Eq.~(\ref{Prepa})]. The extra projective measurement leads to $\rho
_{y}^{se}\rightarrow \Pi _{\alpha }\otimes \mathrm{Tr}_{s}(\Pi _{\alpha
}\rho _{y}^{se})/P(\alpha |y),$ where the conditional probability of $\alpha 
$ given $y$ is $P(\alpha |y)=\mathrm{Tr}_{se}(\Pi _{\alpha }\rho _{y}^{se}).$
The posterior conditional (system) unitary rotation (\ref{Prepa}) leads to $%
\rho _{y}^{se}\rightarrow \mathcal{R}(\alpha |y)[\Pi _{\alpha }]\otimes 
\mathrm{Tr}_{s}(\Pi _{\alpha }\rho _{y}^{se})/P(\alpha |y).$ Thus, \textit{%
in average }(over $\alpha $) the transformation%
\begin{equation}
\rho _{y}^{se}\rightarrow \rho _{y}\otimes \sigma _{e}^{yx},
\end{equation}%
is applied, where the completeness relation $\sum_{\alpha }\Pi _{\alpha }=%
\mathrm{I}_{s}$ was used. As before, the extra quantum operations (\ref%
{Prepa}) are unnecessary for projective $y$-measurements.

The final steps correspond to a bipartite evolution $\mathcal{E}_{\tau }$
during a time interval $\tau =t_{z}-t_{y}$ and the last $z$-measurement,
which leads to $\mathcal{E}_{\tau }[\rho _{y}\otimes \sigma
_{e}^{yx}]\rightarrow \rho _{z}^{se},$ where%
\begin{equation}
\rho _{z}^{se}=\frac{\Omega _{z}\mathcal{E}_{\tau }[\rho _{y}\otimes \sigma
_{e}^{yx}]\Omega _{z}^{\dagger }}{\mathrm{Tr}_{se}(\Omega _{z}^{\dagger
}\Omega _{z}\mathcal{E}_{\tau }[\rho _{y}\otimes \sigma _{e}^{yx}])}.
\label{RhoZAp}
\end{equation}%
The conditional probability of outcome $z$ given that the previous ones were 
$x$ and $y$ is%
\begin{equation}
P(z|y,x)=\mathrm{Tr}_{se}(\Omega _{z}^{\dagger }\Omega _{z}\mathcal{E}_{\tau
}[\rho _{y}\otimes \sigma _{e}^{yx}]).
\end{equation}%
From here and Eq. (\ref{RetroBipartita}), it follows the final expression%
\begin{equation}
P(z,x|y)=\mathrm{Tr}_{se}(\Omega _{z}^{\dagger }\Omega _{z}\mathcal{E}_{\tau
}[\rho _{y}\otimes \sigma _{e}^{yx}])\frac{\mathrm{Tr}_{se}(E_{y}\mathcal{E}%
_{t}[\Omega _{x}\rho _{0}^{se}\Omega _{x}^{\dagger }])}{\sum_{x^{\prime }}%
\mathrm{Tr}_{se}(E_{y}\mathcal{E}_{t}[\Omega _{x^{\prime }}\rho
_{0}^{se}\Omega _{x^{\prime }}^{\dagger }])}.  \label{CPFProbaLast}
\end{equation}

For the calculation of the CPF correlation $C_{pf}(t,\tau
)=\sum_{zx}[P(z,x|y)-P(z|y)P(x|y)]O_{z}O_{x},$ the conditional probability $%
P(z|y)$ can be calculated as $P(z|y)=\sum_{x}P(z,x|y),$ where\ $P(z,x|y)$
follows from Eq. (\ref{CPFProbaLast}). Notice that $P(x|y)=\sum_{z}P(z,x|y),$
by using that $\sum_{z}\Omega _{z}^{\dagger }\Omega _{z}=\mathrm{I}_{s},$
consistently recovers the result (\ref{RetroBipartita}).

\section{Classical environment fluctuations}

Classical environment fluctuations can be described through a stochastic
Liouville equation, $(d/dt)\rho _{t}^{st}=-i\mathcal{L}_{st}(t)[\rho
_{t}^{st}].$ In this situation, the CPF probability follow straightforwardly
from Eq.~(\ref{CPFProbaAp}) after adding the intermediate stochastic
evolution and a corresponding average over realizations (overbar symbol),
which is restricted to the $y$-outcome,%
\begin{equation}
P(z,x|y)=\left. \overline{\mathrm{Tr}(\Omega _{z}^{\dagger }\Omega _{z}%
\mathcal{E}_{t+\tau ,t}^{st}[\rho _{y}])\frac{\mathrm{Tr}(\Omega
_{y}^{\dagger }\Omega _{y}\mathcal{E}_{t,0}^{st}[\Omega _{x}\rho _{0}\Omega
_{x}^{\dagger }])}{\sum_{x^{\prime }}\mathrm{Tr}(\Omega _{y}^{\dagger
}\Omega _{y}\mathcal{E}_{t,0}^{st}[\Omega _{x^{\prime }}\rho _{0}\Omega
_{x^{\prime }}^{\dagger }])}}\right\vert _{y},  \label{PPromedioAPP}
\end{equation}%
where%
\begin{equation}
\mathcal{E}_{t_{b},t_{a}}^{st}\equiv \exp \Big{[}-i\int_{t_{a}}^{t_{b}}dt^{%
\prime }\mathcal{L}_{st}(t^{\prime })\Big{]}.  \label{Etal}
\end{equation}%
When the occurrence of a given $y$-outcome does not depend on the stochastic
evolution in $(0,t)$ the average over realizations in Eq. (\ref{PPromedioAPP}%
) is an unconditional one. Nevertheless, this property is not fulfilled in
general. Thus, for performing the conditional average in Eq. (\ref%
{PPromedioAPP}) is necessary to calculate the probability $P[\mathcal{E}%
_{t,0}^{st}|y]$ of a given (noise) trajectory in $(0,t),$ \textit{labeled}
by the propagator $\mathcal{E}_{t,0}^{st},$ conditioned to the occurrence of
a fixed $y$-outcome. On the other hand, the occurrence of the $y$-outcome is
statistically independent of the noise realizations in the interval $%
(t,t+\tau ).$

From Bayes rule $P[\mathcal{E}_{t,0}^{st}|y]$ can be written as%
\begin{equation}
P[\mathcal{E}_{t,0}^{st}|y]=\frac{P[y|\mathcal{E}_{t,0}^{st}]}{P(y)}P[%
\mathcal{E}_{t,0}^{st}],  \label{PRealization|y}
\end{equation}%
where%
\begin{equation}
P[y|\mathcal{E}_{t,0}^{st}]=\sum_{x}\mathrm{Tr}(\Omega _{y}^{\dagger }\Omega
_{y}\mathcal{E}_{t,0}^{st}[\Omega _{x}\rho _{0}\Omega _{x}^{\dagger }]),
\end{equation}%
while $P[\mathcal{E}_{t,0}^{st}]$ is the probability of each realization.
Thus, Eq. (\ref{PRealization|y}) allows us to write conditional averages in
terms of unconditional averages. Furthermore, from $P[y,\mathcal{E}%
_{t,0}^{st}]=P[y|\mathcal{E}_{t,0}^{st}]P[\mathcal{E}_{t,0}^{st}],$ it
follows%
\begin{equation}
P(y)=\overline{\sum_{x}\mathrm{Tr}(\Omega _{y}^{\dagger }\Omega _{y}\mathcal{%
E}_{t,0}^{st}[\Omega _{x}\rho _{0}\Omega _{x}^{\dagger }])},
\end{equation}%
where here the overbar denotes an unconditional average over realizations.
Introducing Eq. (\ref{PRealization|y}) into Eq. (\ref{PPromedioAPP}) we get
the final expression%
\begin{equation}
P(z,x|y)=\overline{\frac{\mathrm{Tr}(\Omega _{z}^{\dagger }\Omega _{z}%
\mathcal{E}_{t+\tau ,t}^{st}[\rho _{y}])\mathrm{Tr}(E_{y}\mathcal{E}%
_{t,0}^{st}[\Omega _{x}\rho _{0}\Omega _{x}^{\dagger }])}{\overline{%
\sum_{x^{\prime }}\mathrm{Tr}(E_{y}\mathcal{E}_{t,0}^{st}[\Omega _{x^{\prime
}}\rho _{0}\Omega _{x^{\prime }}^{\dagger }])}}},  \label{PZXStochastic}
\end{equation}%
which is written in terms of unconditional averages (overbars symbols). The
previous expression can be rewritten with a similar structure to that of
quantum environments [Eq. (\ref{CPFProbaLast})]%
\begin{equation}
P(z,x|y)=\overline{\mathrm{Tr}(\Omega _{z}^{\dagger }\Omega _{z}\mathcal{E}%
_{t+\tau ,t}^{st}[\rho _{yx}^{st}])}\frac{\overline{\mathrm{Tr}(E_{y}%
\mathcal{E}_{t,0}^{st}[\Omega _{x}\rho _{0}\Omega _{x}^{\dagger }])}}{%
\overline{\sum_{x^{\prime }}\mathrm{Tr}(E_{y}\mathcal{E}_{t,0}^{st}[\Omega
_{x^{\prime }}\rho _{0}\Omega _{x^{\prime }}^{\dagger }])}},
\label{STCHalaBath}
\end{equation}%
where%
\begin{equation}
\rho _{yx}^{st}\equiv \rho _{y}\frac{\mathrm{Tr}(E_{y}\mathcal{E}%
_{t,0}^{st}[\Omega _{x}\rho _{0}\Omega _{x}^{\dagger }])}{\overline{\mathrm{%
Tr}(E_{y}\mathcal{E}_{t,0}^{st}[\Omega _{x}\rho _{0}\Omega _{x}^{\dagger }])}%
}.  \label{NoiseCorrelatedState}
\end{equation}

\subsection{Markovianity of white fluctuations}

The system dynamics induced by the stochastic Liouville equation is
Markovian if $P(z,x|y)=P(z|y)P(x|y)$ $[C_{pf}(t,\tau )=0].$ From Eq. (\ref%
{PZXStochastic}) [or (\ref{STCHalaBath})] it is simple to derive the
condition%
\begin{equation}
\overline{\mathrm{Tr}(\Omega _{z}^{\dagger }\Omega _{z}\mathcal{E}_{t+\tau
,t}^{st}[\rho _{y}])\mathrm{Tr}(\Omega _{y}^{\dagger }\Omega _{y}\mathcal{E}%
_{t,0}^{st}[\Omega _{x}\rho _{0}\Omega _{x}^{\dagger }])}=\overline{\mathrm{%
Tr}(\Omega _{z}^{\dagger }\Omega _{z}\mathcal{E}_{t+\tau ,t}^{st}[\rho _{y}])%
}\times \overline{\mathrm{Tr}(\Omega _{y}^{\dagger }\Omega _{y}\mathcal{E}%
_{t,0}^{st}[\Omega _{x}\rho _{0}\Omega _{x}^{\dagger }])}.
\label{PPromedioAP}
\end{equation}%
White fluctuations guarantee this property. The demonstration is given below.

By an adequate change of system operators base one can always write $%
\mathcal{L}_{st}(t)=\sum_{j}\xi _{j}(t)\mathcal{L}_{j},$ where $\{\xi
_{j}(t)\}$ are independent white noises and $\mathcal{L}_{j}$ are
deterministic superoperators. Using that Eq. (\ref{PPromedioAP}) is defined
by the product of two real terms (probabilities), the dynamics is Markovian
if for each noise $\xi _{j}(t)\rightarrow \xi (t)$ it is valid that 
\begin{equation}
\overline{\Xi (f(t+\tau ,t))\ \Xi ^{\prime }(f(t,0))}=\overline{\Xi
(f(t+\tau ,t))}\times \overline{\Xi ^{\prime }(f(t,0))},
\end{equation}%
where here $f(t_{b},t_{a})\equiv \mathrm{Re}[\exp
-i\int_{t_{a}}^{t_{b}}dt^{\prime }\xi (t^{\prime })].$ The functions $\Xi
(x) $ and $\Xi ^{\prime }(x)$ depend on the underlying superoperators. Given
that they admit a series expansion, it follows the equivalent condition $%
[\Xi (x)\rightarrow x,$ $\Xi ^{\prime }(x)\rightarrow x]$%
\begin{equation}
\overline{f(t+\tau ,t)\ f(t,0)}=\overline{f(t+\tau ,t)}\times \overline{%
f(t,0)}.
\end{equation}%
This equation is fulfilled by arbitrary white noises. In fact, the real
parts can be written as $\mathrm{Re}[e^{-i\Phi }]=(e^{+i\Phi }+e^{-i\Phi
})/2,$ leading to the condition%
\begin{equation}
\overline{\exp [i(\Phi _{t+\tau ,t}\pm \Phi _{t,0})]}=\overline{\exp (i\Phi
_{t+\tau ,t})}\times \overline{\exp (\pm i\Phi _{t,0})},
\label{ConditionAPP}
\end{equation}%
where $\Phi _{t_{b},t_{a}}\equiv \int_{t_{a}}^{t_{b}}dt^{\prime }\xi
(t^{\prime }).$ These averages can be performed in an exact way by
introducing the noise characteristic functional \cite{vanKampen} $G[k]\equiv 
\overline{\exp i\int_{0}^{\infty }dt^{\prime }k(t)\xi (t^{\prime })]}.$ From
a series expansion in cumulants, for an arbitrary white noise it reads%
\begin{equation}
G[k]=\exp \sum_{m=1}^{\infty }\frac{i^{m}}{m!}\Gamma _{m}\int_{0}^{\infty
}dt^{\prime }(k(t^{\prime }))^{m},
\end{equation}%
where $\Gamma _{m}$ measures the $m$-cumulant of the noise. Taking the test
function $k(t)=\theta (t+\tau -t^{\prime })\theta (t^{\prime }-t)\pm \theta
(t-t^{\prime }),$ where $\theta (x)$ is the step function, condition (\ref%
{ConditionAPP}) follows straightforwardly after simple calculation steps.
Gaussian white noises are defined by the condition $\Gamma _{m}=0$ if $m>2.$

\section{Higher order conditional past-future correlations}

Given a sequence of time ordered random events $x\rightarrow
y_{1}\rightarrow \cdots y_{n}\rightarrow z$ occurring at times $%
t_{x}<t_{y_{1}}\cdots <t_{y_{n}}<t_{z},$ classical Markovianity implies the
conditions \cite{vanKampen} $P(z|y_{1},x)=P(z|y_{1}),$ and for arbitrary $n,$
$P(z|y_{n}\cdots y_{1},x)=P(z|y_{n}).$ These constraints can be re-expressed
in terms of a \textit{hierarchical set of CPF correlations} as follows. From
Bayes rule, the conditional probability $P(z,x|y_{n}\cdots y_{1})$ of past
and future events conditioned to a set of $n$-intermediate states can be
written as%
\begin{equation}
P(z,x|y_{n},\cdots y_{1})=P(z|y_{n},\cdots y_{1},x)P(x|y_{n},\cdots y_{1}).
\end{equation}%
A $n$-order CPF correlation is defined as%
\begin{equation}
C_{pf}^{(n)}=\sum_{zx}[P(z,x|y_{n},\cdots y_{1})-P(z|y_{n},\cdots
y_{1})P(x|y_{n},\cdots y_{1})]O_{z}O_{x},  \label{nCPF}
\end{equation}%
where $O_{z}$ and $O_{x}$ are system observables. A $n$\textit{-order CPF
independence} is defined by the conditions $C_{pf}^{(k)}=0$ $\forall k\leq
n, $ and $C_{pf}^{(n+1)}\lessgtr 0.$ These conditions are fulfilled when $%
P(z|y_{k}\cdots y_{1},x)=P(z|y_{k}\cdots y_{1})$ $\forall k\leq n,$ and $%
P(z|y_{n+1}\cdots y_{1},x)\neq P(z|y_{n+1}\cdots y_{1}).$ Consequently,
classical Markovianity can alternatively be defined by the validity of CPF
independence to any order, that is, $C_{pf}^{(n)}=0$ $\forall n.$ Below $%
C_{pf}^{(n)}$ is extended to a quantum regime.

\subsection{Markovianity of quantum measurements}

The (first order) CPF independence fulfilled by a set of three consecutive
quantum measurements [Eq. (\ref{CPFProbaAp})] can be extended to an
arbitrary order. The sequence of random results $x\rightarrow
y_{1}\rightarrow \cdots y_{n}\rightarrow z$ is related to a set of
measurement operators $\Omega _{x},$ $\Omega _{y_{1}},$ $\cdots ,$ $\Omega
_{y_{n}},$ and $\Omega _{z},$ which obey $\sum_{x}\Omega _{x}^{\dagger
}\Omega _{x}=\sum_{y_{1}}\Omega _{y_{1}}^{\dagger }\Omega _{y_{1}}\cdots
=\sum_{y_{n}}\Omega _{y_{n}}^{\dagger }\Omega _{y_{n}}=\sum_{z}\Omega
_{z}^{\dagger }\Omega _{z}=\mathrm{I}.$ The sum indexes run over all
possible outcomes at each stage. Furthermore, the $y_{n}$-measurement is
taken as a projective one, or equivalently a causal break or preparation is
performed after it [Eq.~(\ref{Prepa})]. The associated system state is $\rho
_{y_{n}}.$ The remaining measurements are arbitrary. By calculating the
system state after each measurement event, the generalization of Eq.~(\ref%
{CPFProbaAp}) reads%
\begin{equation}
P(z,x|y_{n},\cdots y_{1})=\mathrm{Tr}(\Omega _{z}^{\dagger }\Omega _{z}[\rho
_{y_{n}}])\frac{\mathrm{Tr}(E_{y}^{(n)}[\Omega _{x}\rho _{0}\Omega
_{x}^{\dagger }])}{\sum_{x^{\prime }}\mathrm{Tr}(E_{y}^{(n)}[\Omega
_{x^{\prime }}\rho _{0}\Omega _{x^{\prime }}^{\dagger }])},
\label{PzxNMeasurement}
\end{equation}%
where the first and second factors correspond to $P(z|y_{n}\cdots y_{1},x)$\
and $P(x|y_{n}\cdots y_{1})$\ respectively. Furthermore, the effect operator 
$E_{y}^{(n)}$ is%
\begin{equation}
E_{y}^{(n)}\equiv \Omega _{y_{1}}^{\dagger }\cdots \Omega _{y_{n}}^{\dagger
}\Omega _{y_{n}}\cdots \Omega _{y_{1}}.
\end{equation}%
Eq. (\ref{PzxNMeasurement}) say us that quantum measurements fulfill CPF
independence to any order. In fact, $P(z|y_{n}\cdots y_{1},x)=P(z|y_{n}).$
Notice that Eq. (\ref{PzxNMeasurement}) can be read from Eq. (\ref%
{CPFProbaAp}) under the replacement $E_{y}\rightarrow E_{y}^{(n)}$ of the
effect operator.

\subsection{n-degree quantum Markovian evolutions}

By adding the system evolution between measurements, Eq.~(\ref%
{PzxNMeasurement}) becomes%
\begin{equation}
P(z,x|y_{n},\cdots y_{1})=\mathrm{Tr}(\Omega _{z}^{\dagger }\Omega _{z}%
\mathcal{E}^{\prime }[\rho _{y_{n}}])\frac{\mathrm{Tr}(E_{y}^{(n)}\mathcal{E}%
[\Omega _{x}\rho _{0}\Omega _{x}^{\dagger }])}{\sum_{x^{\prime }}\mathrm{Tr}%
(E_{y}^{(n)}\mathcal{E}[\Omega _{x^{\prime }}\rho _{0}\Omega _{x^{\prime
}}^{\dagger }])}.  \label{nMarkov}
\end{equation}%
Here, $\mathcal{E}^{\prime }=\mathcal{E}_{t_{z}t_{y_{n}}}$ and $\mathcal{E}=%
\mathcal{E}_{t_{y_{1}}t_{x}},$ where $\mathcal{E}_{t_{b}t_{a}}$ is the\
system density matrix propagator between the times $t_{a}$ and $t_{b}.$
Furthermore, the effect operator reads%
\begin{equation}
E_{y}^{(n)}=\Omega _{y_{1}}^{\dagger }\mathcal{E}_{t_{y_{2}}t_{y_{1}}}^{\#}[%
\Omega _{y_{2}}^{\dagger }\cdots \mathcal{E}_{t_{y_{n-1}}t_{y_{n-2}}}^{\#}[%
\Omega _{y_{n-1}}^{\dagger }\mathcal{E}_{t_{y_{n}}t_{y_{n-1}}}^{\#}[\Omega
_{y_{n}}^{\dagger }\Omega _{y_{n}}]\Omega _{y_{n-1}}]\cdots \Omega
_{y_{2}}]\Omega _{y_{1}}.  \label{Eyn}
\end{equation}%
The system dual propagator is defined by the equality $\mathrm{Tr}(\Omega 
\mathcal{E}[\rho ])=\mathrm{Tr}(\rho \mathcal{E}^{\#}[\Omega ]),$ where $%
\rho $ and $\Omega $ are arbitrary system operators. As usual, the effect
operator $E_{y}^{(n)}$ [Eq.~(\ref{Eyn})] \textquotedblleft
evolves\textquotedblright\ in a time reversed order \cite{molmer}.

The system evolution is defined as Markovian of degree $n$ when it does not
break CPF independence up to order $n.$ From Eq. (\ref{nMarkov}) if follows
that this condition is fulfilled when the system propagator $\mathcal{E}%
_{t_{z}t_{y_{n}}}$\ does not depend on the past measurement outcomes
occurring at times $t_{x}<t_{y_{1}}\cdots <t_{y_{n-1}}.$ In particular, a
deterministic unitary dynamics is Markovian at all orders. The same property
is fulfilled when a Born-Markov approximation applies.

\subsection{Quantum system-environment models}

The conditional probability $P(z,x|y_{n},\cdots y_{1})$ can explicitly be
calculated for bipartite system-environment models. The generalization of
Eq. (\ref{CPFProbaLast}) reads 
\begin{equation}
P(z,x|y_{n},\cdots y_{1})=\mathrm{Tr}_{se}(\Omega _{z}^{\dagger }\Omega _{z}%
\mathcal{E}_{\tau }[\rho _{y_{n}}\otimes \sigma _{e}^{y_{n}\cdots y_{1},x}])%
\frac{\mathrm{Tr}_{se}(E_{y}^{(n)}\mathcal{E}_{t}[\Omega _{x}\rho
_{0}^{se}\Omega _{x}^{\dagger }])}{\sum_{x^{\prime }}\mathrm{Tr}%
_{se}(E_{y}^{(n)}\mathcal{E}_{t}[\Omega _{x^{\prime }}\rho _{0}^{se}\Omega
_{x^{\prime }}^{\dagger }])},  \label{n-Bipartito}
\end{equation}%
where $\tau \equiv t_{z}-t_{y_{n}}$ and $t\equiv t_{y_{1}}-t_{x}.$ $\mathcal{%
E}_{t}$ is the\ bipartite system-environment propagator. The environment
state $\sigma _{e}^{y_{n}\cdots y_{1},x}$ is%
\begin{equation}
\sigma _{e}^{y_{n}\cdots y_{1},x}=\frac{\mathrm{Tr}_{e}(E_{y}^{(n)}\mathcal{E%
}_{t}[\Omega _{x}\rho _{0}^{se}\Omega _{x}^{\dagger }])}{\mathrm{Tr}%
_{se}(E_{y}^{(n)}\mathcal{E}_{t}[\Omega _{x}\rho _{0}^{se}\Omega
_{x}^{\dagger }])},
\end{equation}%
while the effect operator $E_{y}^{(n)}$ here reads%
\begin{equation}
E_{y}^{(n)}=\Omega _{y_{1}}^{\dagger }\Omega _{y_{2}}^{\dagger
}(t_{y_{2}},t_{y_{1}})\cdots \Omega _{y_{n}}^{\dagger
}(t_{y_{n}},t_{y_{1}})\Omega _{y_{n}}(t_{y_{n}},t_{y_{1}})\cdots \Omega
_{y_{2}}(t_{y_{2}},t_{y_{1}})\Omega _{y_{1}},  \label{EfectoNDynamico}
\end{equation}%
where%
\begin{equation}
\Omega _{y}(t_{b},t_{a})=\mathcal{E}_{t_{b}-t_{a}}^{\#}[\Omega
_{y}]=e^{+iH_{T}(t_{b}-t_{a})}\Omega _{y}e^{-iH_{T}(t_{b}-t_{a})}.
\end{equation}%
$H_{T}$ is the total system-environment Hamiltonian. Notice that Eq. (\ref%
{n-Bipartito}) can be read from Eq. (\ref{CPFProbaLast}) under the
replacement $E_{y}\rightarrow E_{y}^{(n)}.$

\subsection{Classical environment fluctuations}

For systems driven by classical noise fluctuations, the generalization of
Eq.~(\ref{PZXStochastic}) reads 
\begin{equation}
P(z,x|y_{n},\cdots y_{1})=\frac{\overline{\mathrm{Tr}(\Omega _{z}^{\dagger
}\Omega _{z}\mathcal{E}_{t_{y_{n}}+\tau ,t_{y_{n}}}^{st}[\rho _{y_{n}}])%
\mathrm{Tr}(E_{y}^{(n)}\mathcal{E}_{t,0}^{st}[\Omega _{x}\rho
_{0}^{se}\Omega _{x}^{\dagger }])}}{\overline{\sum_{x^{\prime }}\mathrm{Tr}%
(E_{y}^{(n)}\mathcal{E}_{t,0}^{st}[\Omega _{x^{\prime }}\rho _{0}^{se}\Omega
_{x^{\prime }}^{\dagger }])}},  \label{N-Stoch}
\end{equation}%
where $\tau =t_{z}-t_{y_{n}}$ and $t=t_{y_{1}}-t_{x}.\ $Similarly to Eq. (%
\ref{STCHalaBath}), this expression can be rewritten with the structure%
\begin{equation}
P(z,x|y_{n},\cdots y_{1})=\overline{\mathrm{Tr}(\Omega _{z}^{\dagger }\Omega
_{z}\mathcal{E}_{t_{y_{n}}+\tau ,t_{y_{n}}}^{st}[\rho _{y_{n}\cdots
y_{1},x}^{st}])}\frac{\overline{\mathrm{Tr}(E_{y}^{(n)}\mathcal{E}%
_{t,0}^{st}[\Omega _{x}\rho _{0}^{se}\Omega _{x}^{\dagger }])}}{\overline{%
\sum_{x^{\prime }}\mathrm{Tr}(E_{y}^{(n)}\mathcal{E}_{t,0}^{st}[\Omega
_{x^{\prime }}\rho _{0}^{se}\Omega _{x^{\prime }}^{\dagger }])}},
\label{N-StochSecond}
\end{equation}%
where%
\begin{equation}
\rho _{y_{n}\cdots y_{1},x}^{st}=\rho _{y}\frac{\mathrm{Tr}(E_{y}^{(n)}%
\mathcal{E}_{t,0}^{st}[\Omega _{x}\rho _{0}\Omega _{x}^{\dagger }])}{%
\overline{\mathrm{Tr}(E_{y}^{(n)}\mathcal{E}_{t,0}^{st}[\Omega _{x}\rho
_{0}\Omega _{x}^{\dagger }])}}.
\end{equation}%
In the previous expressions, the effect operator $E_{y}^{(n)}$ is given by
Eq.~(\ref{EfectoNDynamico}) where now%
\begin{equation}
\Omega _{y}(t_{b},t_{a})=(\mathcal{E}_{t_{b}t_{a}}^{st})^{\#}[\Omega
_{y}]=\exp \left[ +i\int_{t_{a}}^{t_{b}}dt^{\prime }\mathcal{L}%
_{st}^{\#}(t^{\prime })\right] \Omega _{y}.
\end{equation}%
Notice that Eqs. (\ref{N-Stoch}) and (\ref{N-StochSecond}) can be read
respectively from Eqs. (\ref{PZXStochastic}) and (\ref{STCHalaBath}) under
the replacement $E_{y}\rightarrow E_{y}^{(n)}.$

\section{Minimal measurement set for detecting memory effects in open
quantum systems}

Expressions (\ref{n-Bipartito}) and (\ref{N-StochSecond}) allow us to define
higher order CPF correlations [Eq. (\ref{nCPF})] for open quantum systems.
Nevertheless, in contrast with classical systems, conditions such as $%
C_{pf}^{(k)}=0$ $\forall k\leq n$ and $C_{pf}^{(n+1)}\lessgtr 0$ are not
expected. In fact, due to the degree of freedom given by the measurement
operators, for standard open quantum systems it is expected $%
C_{pf}^{(1)}\lessgtr 0.$ Thus, memory effects can be analyzed over the basis
of a minimal three quantum-measurements scheme associated to the first order
CPF\ correlation.

The property $C_{pf}^{(1)}\lessgtr 0$ can be derived by studying the
conditions under which it vanishes. For \textit{quantum system-environment
models} [Eq. (\ref{CPFProbaLast})], $C_{pf}^{(1)}=0$ when the environment
state $\sigma _{e}^{yx}$ [Eq.~(\ref{BathStateYX})] does not depend on $x$%
-outcomes, $\sigma _{e}^{yx}=\sigma _{e}^{y}.$ This property must be valid
for arbitrary measurement operators $\{\Omega _{x}\}$ and $\{\Omega _{y}\}.$ 
\textit{This extra degree of freedom is absent in the classical domain.} The
standard Born-Markov approximation guarantees the fulfillment of this
condition. Beyond this approximation the system dynamics breaks CPF
independence $[C_{pf}^{(1)}\lessgtr 0].$ In fact, assuming for example an
uncorrelated initial state $\rho _{0}^{se}=\rho _{0}\otimes \sigma _{e},$
with projective measurement operators $\{\Omega _{x}=|x\rangle \langle x|\}$
and $\{\Omega _{y}=|y\rangle \langle y|\},$ Eq. (\ref{BathStateYX}) becomes%
\begin{equation}
\sigma _{e}^{yx}=\frac{\langle y|\mathcal{U}_{t}|x\rangle \sigma _{e}\langle
x|\mathcal{U}_{t}^{\dag }|y\rangle }{\mathrm{Tr}_{e}[\langle x|\mathcal{U}%
_{t}^{\dag }|y\rangle \langle y|\mathcal{U}_{t}|x\rangle \sigma _{e}]},
\end{equation}%
where $\mathcal{U}_{t}\equiv \exp [-itH_{T}].$ CPF independence $%
[C_{pf}^{(1)}=0]$\ is valid if \textit{this environment state is independent
of the arbitrary system state }$|x\rangle $\textit{\ for any arbitrary
system state }$|y\rangle .$ This property is fulfilled when $%
H_{T}=H_{s}+H_{e}.$ Thus, the system and the environment, with Hamiltonians $%
H_{s}$ and $H_{e}$ respectively, do not interact (closed system). In this
case, all higher correlations also vanish, $C_{pf}^{(n)}=0.$ Consequently,
for open quantum systems interaction with the environment leads in general
to $C_{pf}^{(1)}\lessgtr 0.$ Correlated system-environment initial
conditions do not change this result. Based on Eqs.~(\ref{STCHalaBath}) and (%
\ref{NoiseCorrelatedState}) a similar conclusion is also valid for \textit{%
quantum system coupled to standard noise sources}. On the other hand, from a
formal point of view, the conditions $C_{pf}^{(k)}=0$ $\forall k\leq n$ and $%
C_{pf}^{(n+1)}\lessgtr 0$ may be fulfilled by an open quantum system coupled
to a classical noise source (like in a discrete-time collisional model) that
by itself satisfy these conditions ($n$-order CPF independence). Standard
noises such as non-white Gaussian or dichotomic fluctuations do not fulfill
these constraints.
\end{widetext}


\begin{thebibliography}{99}
\bibitem{comment1} In order to avoid confusion with the measurements
notation: $x,$ $y,$ $z,$ directions in the Bloch sphere are denoted with a
hat symbol: $\hat{x},$ $\hat{y},$ $\hat{z}.$
\end{thebibliography}
\end{document}